\documentclass[12pt,twocolumn,twoside]{article}   

\usepackage[super,sort,comma]{natbib}

\usepackage{booktabs}

\usepackage{fancyhdr}		
\usepackage[section]{placeins}   %

\usepackage{graphicx}

\makeatletter \renewcommand\@biblabel[1]{$^{#1}$} \makeatother
\usepackage[hidelinks]{hyperref}
\hypersetup{
	colorlinks=true,
	breaklinks=true,
	citecolor=blue,
    filecolor=blue,
	urlcolor=blue,
    linkcolor=blue,
	pdfpagemode=UseNone,
	bookmarksopen=false, 
	pdftitle={Predicting successful clinical candidates for fiducial-free lung tumor tracking with a deep learning binary classification model},
	pdfauthor={Matthieu Lafreniere},
	pdfkeywords={Artificial Intelligence, Deep Learning, Transfer Learning, Binary Classification, LOT Simulation, CyberKnife, Accuray, Respiratory Motion Modeling, Respiratory Motion, Motion Modeling, Lung Tumors, Lung Cancer, Radiation Therapy, SBRT, 4DCT, Digitally Reconstructed Radiograph (DRR)}
}  

\newcommand\figureWidth{1.00}

\usepackage{xcolor}
\definecolor{gray}{rgb}{0.6,0.6,0.6}
\definecolor{red}{rgb}{0.85,0,0}
\definecolor{green}{rgb}{0,0.85,0}
\definecolor{blue}{rgb}{0,0,0.85}
\definecolor{beige}{rgb}{0.92,0.87,0.78}

\usepackage[all]{hypcap}    

\begin{document}

\onecolumn

\pagenumbering{roman}
\setcounter{page}{1}
\pagestyle{plain}

\center{\sf {\Large {\bfseries Predicting successful clinical candidates for fiducial-free lung tumor tracking with a deep learning binary classification model} \\  
\vspace*{5mm}
{\small In loving memory of my Grand-Father Denis.}\\
\vspace*{5mm}
M. Lafreni\`ere, G. Valdes, M. Descovich} \\
University of California, San Francisco, 1600 Divisadero Street, Suite H1031, San Francisco, CA 94143-1708, USA
}


E-mail addresses of authors to whom correspondence should be addressed: matthieulafreniere@hotmail.com, martina.descovich@ucsf.edu

\begin{abstract}

\noindent 

{\bf Objectives:} The CyberKnife system is a robotic radiosurgery platform that allows the delivery of lung SBRT treatments using fiducial-free soft-tissue tracking. However, not all lung cancer patients are eligible for lung tumor tracking. Tumor size, density and location impact the ability to successfully detect and track a lung lesion in 2D orthogonal X-ray images. The standard workflow to identify successful candidates for lung tumor tracking is called Lung Optimized Treatment (LOT) simulation, and involves multiple steps from CT acquisition to the execution of the simulation plan on CyberKnife. The aim of the study is to develop a deep learning classification model to predict which patients can be successfully treated with lung tumor tracking, thus circumventing the LOT simulation process.\\

{\bf Methods:} Target tracking is achieved by matching orthogonal x-ray images with a library of digital radiographs reconstructed from the simulation CT scan (DRRs). We developed a deep learning model to create a binary classification of lung lesions as being trackable or untrackable based on tumor template DRR extracted from the CyberKnife system, and tested five different network architectures. The study included a total of 271 images (230 trackable, 41 untrackable) from 129 patients with one or multiple lung lesions. 80\% of the images were used for training, 10\% for validation, and the remaining 10\% for testing.\\

{\bf Results:} For all 5 convolutional neural networks, the binary classification accuracy reached 100\% after training, both in the validation and the test set, without any false classifications.\\

{\bf Conclusions:} A deep learning model can distinguish features of trackable and untrackable lesions in DRR images, and can predict successful candidates for fiducial-free lung tumor tracking.\\

\end{abstract}

\twocolumn


\tableofcontents

\newpage

\setlength{\baselineskip}{0.7cm}      

\pagenumbering{arabic}
\setcounter{page}{1}
\pagestyle{fancy}
\raggedright

\section{Introduction}

    The CyberKnife robotic radiosurgery system (Accuray Inc., CA, USA) is a medical device that allows the treatment of lung lesions using stereotactic localization and real-time motion tracking \cite{Kilby2020}. Stereotactic localization is achieved by automatic registration of two orthogonal X-ray images at 45 degrees oblique views (image A and image B) to pre-generated DRRs obtained from the patient CT scan. The target can be localized based on density difference between the tumor and the surrounding lung tissue (lung tumor tracking) or by implanting fiducial markers in the proximity of the tumor (fiducial tracking). Fiducial tracking and lung tumor tracking can be used in conjunction with Synchrony motion modeling to dynamically track the position of the tumor and re-direct the radiation beams to the continuously changing target location. If a lung lesion is visible on both projection images (2-views), the target motion is tracked in 3 dimensions \cite{Yang2017TargetMD, Bahig2013}. If a lung tumor is visible only on one image (1-view), the target motion parallel to the imaging plane is tracked, and additional margins are included to compensate for out-of-plane motion. If direct tumor localization is not possible, the bony anatomy of nearby vertebral bodies can be used for alignment \cite{BAUMANN2018}. Spine tracking (also called 0-view) does not allow to account for target motion and requires the inclusion of a large treatment volume encompassing all possible tumor positions throughout the respiratory cycle. To minimize the possibility of mismatch between anatomy-based registration and actual tumor position, spine tracking should only be used for lesions that are in close proximity to the spine and present relative small motion \cite{GUO2015}. 

    Lung tumor tracking is clinically desirable to keep target margins to a minimum, while avoiding the risks associated with the fiducial implantation procedure \cite{Bhagat2010, Loo2011}. However, not all lung tumors can be directly visualized on planar radiographs, and prospectively identifying candidates for fiducial-free lung tumor tracking is challenging.

	\begin{figure*}[!ht]
		\centering 
		\includegraphics[width=\figureWidth\linewidth]
		{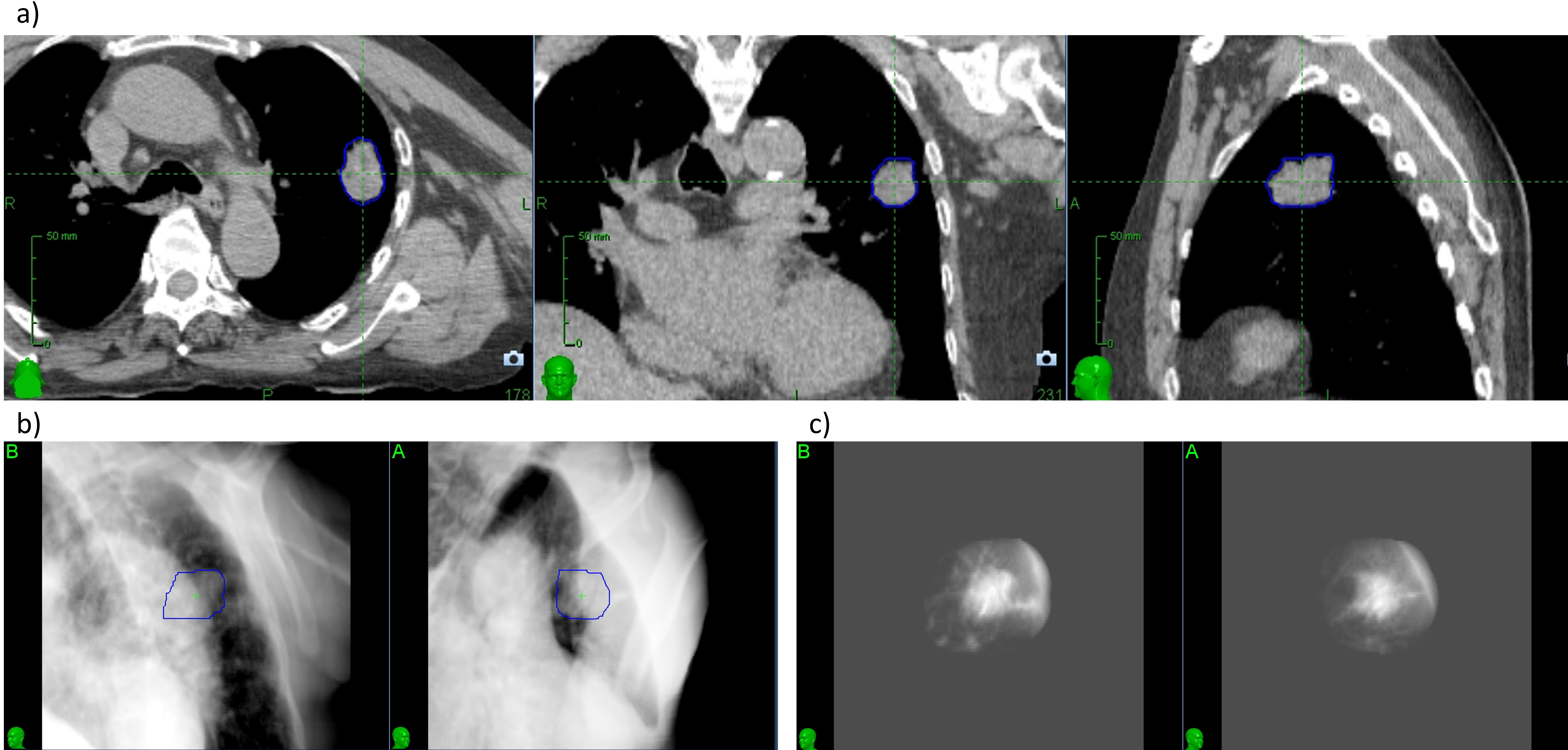}
		\caption[]{{Example of a lesion that can be tracked with 2-views lung tumor tracking: a) axial, coronal and sagittal CT slices showing the lesion; b) full content DRR; c) Tumor template DRR.}}   
		\label{XLT_patient_example}
	\end{figure*}
	
	\begin{figure}[!ht]
		\centering 
		\includegraphics[width=\figureWidth\linewidth]
		{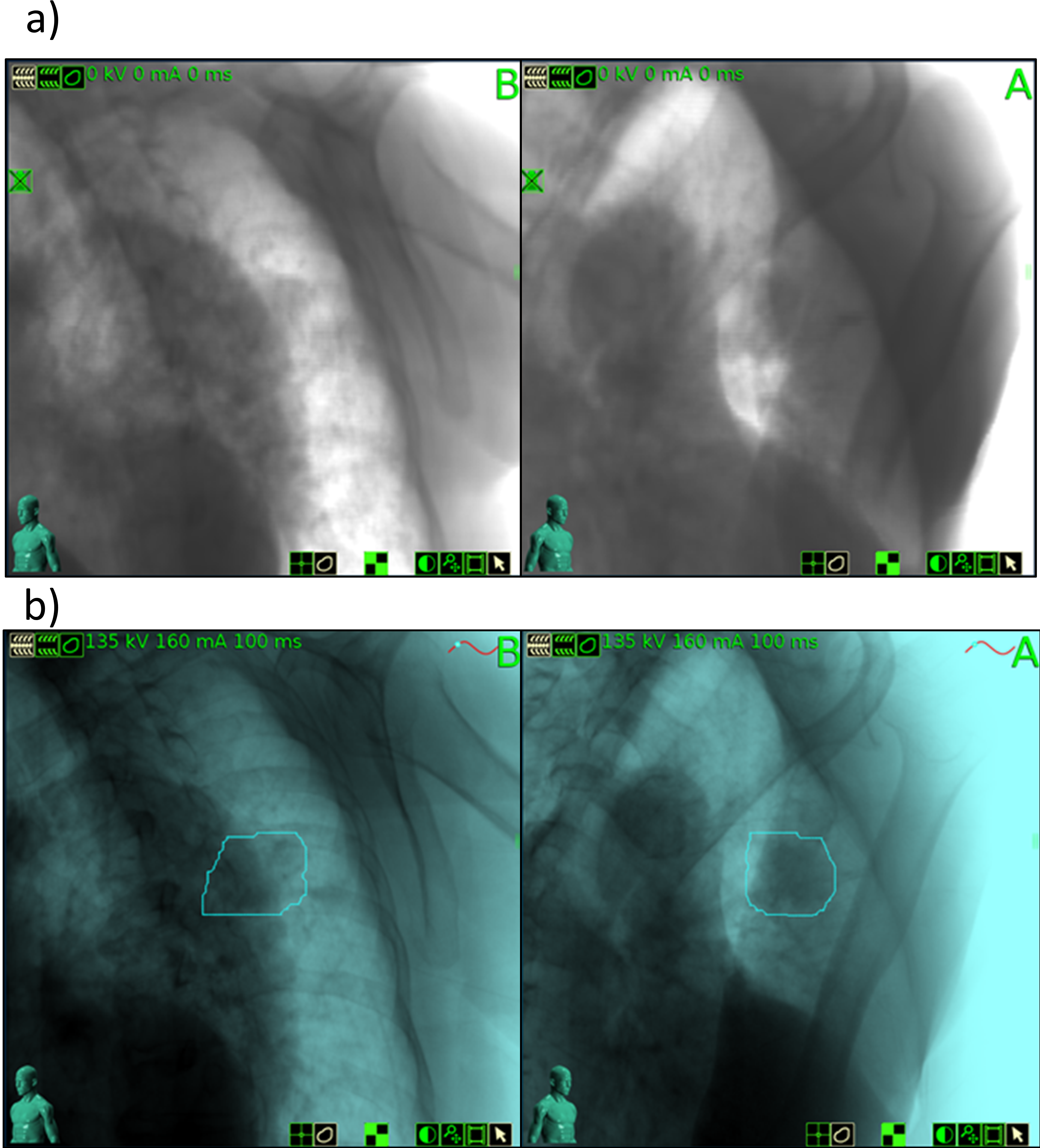}
		\caption[]{{Treatment images for the same lesion displayed in figure 1: a) DRR; b) DRR overlay with 135 kV X-ray image.}}   
		\label{Treatment_Images}
	\end{figure}
 
    The visibility of a lung lesion on 2D projection images depends on its size, density, shape and position relative to the spine or mediastinum. Typically, lung tumor tracking is successful for lesions greater than 1.0-1.5 cm in all dimensions, located in the peripheral region of the lung, and not obstructed by the spine or the heart in the 45 degree view. Figure \ref{XLT_patient_example}a shows axial, sagittal and coronal CT slices of a patient with an upper left lung lesion (2 cm x 2 cm x 3 cm) that was treated using lung tumor tracking. Figure \ref{XLT_patient_example}b shows the full content DRR generated from the planning CT, and figure \ref{XLT_patient_example}c shows the tumor template DRR, with the background removed. The blue contour around the lesion represents the target tracking volume (TTV). The TTV is projected on the DRR to perform the registration based on the tumor intensity, and it is used to visually assess proper target identification. The tumor template DRR is generated by limiting the ray-tracing projections through a volume of the planning CT around the TTV, thus enhancing the visibility of the tumor, and the ability of the tracking algorithm to identify its location. Registration between tumor template DRR and X-ray image is performed separately for each view, therefore image A and image B can be analyzed independently. Figure \ref{Treatment_Images} shows, for the same patient, the DRR visible at the treatment console (a) and the overlay between the DRR and the X-ray images (b). 
    
    Assessing whether a given patient is a suitable candidate for lung tumor tracking is a resource-intensive and time-consuming process, even though it is needed only for critical cases. This process - called lung optimized tracking (LOT) simulation -  takes approximately one hour, and requires multiple steps: 1) acquisition of a 4DCT scan, or inhale and exhale breath-hold scans to avoid motion artifacts and target blurring; 2) generation of a simulation plan on the treatment planning system; 3) execution of the simulation plan on the treatment delivery platform. At each respiratory phase, the tracking system calculates the location of the target, and the physician validates the lesion identification on image A and image B, independently. If the tracking system correctly identifies the lesion location in more than 75\% of the respiratory phases, then lung tumor tracking is deemed clinically safe. 
	 
    The goal of this study was to develop an automation tool for determining a priory the optimal lung tumor tracking method. Leveraging the computing power of Artificial Intelligence (AI), we opted for a deep learning approach to create a binary classification of lung cancer lesions as being trackable or untrackable based on tumor template DRRs sourced from simulation CT scans. This novel deep-learning based binary classification of lung tumor trackability technique allows to circumvent the LOT simulation process altogether, and will help physicians in selecting the best treatment option for their patients.
    
    Deep learning is a subcategory of machine learning based on several layers of artificial neural networks that extract increasingly more abstract features as the analyzed data is processed from superficial layers into deeper layers. Transfer learning is a type of deep learning that allows to retain knowledge obtained by using a neural network that was trained on a large data set, and then retraining the network on a different and smaller data set, while reducing the training weights in the early layers and focusing training in the last layers. Transfer learning increases learning speed and is useful when working with a small data set, which is why we opted for this approach.


\section{Methods}

\subsection{Deep Learning Technique Overview}
	
    This study was conducted under IRB approval and included 129 de-identified patients with a total 144 lesions (115 trackable and 29 untrackable). The patients were treated at our institution between 2011 and 2017. The images obtained retrospectively from actual patient treatments were labeled as trackable if the lung tumor tracking system was able to track the tumors during treatment. The images were labeled as untrackable if the lesions were not tracked during treatment, either because the result of the LOT simulation was 0-view tracking, or because clinical experts (physician and physicist) reviewed the DRR and determined that lung tumor tracking would not be successful. This combination of knowledge concerning tumor trackability is important in order to build a good deep learning model based on a valid ground truth that can produce an accurate representation of reality. 
    
    For each lesion, there exist 2 images at 2 different projection angles (image A and image B). In the context of training our deep learning algorithm, image A and image B are considered independent from each other since the objective is to achieve binary classification with only two possible outcomes: identification of tumors that are trackable and tumors that are untrackable. For trackable lesions, the tumor template DRR used during treatment were extracted from CyberKnife. For untrackable lesions, tumor template DRRs had to be generated specifically for this study. For these 29 lesions, dummy lung tumor tracking plans were created. Plans were saved as deliverable, and DRRs were generated by CyberKnife, extracted at the delivery console and reviewed by an expert. For 12 lesions, both images were deemed not adequate for lung tumor tracking, and were labeled as untrackable. For 17 lesions, one image was labeled as untrackable, and one image was questionable (i.e. clinical experts could not rule out the possibility that that lesion could have been potentially treated with 1-view tracking). Questionable images were excluded from the analysis. The study included a total of 271 images: 230 images of trackable lesions and 41 images of untrackable lesions. The process of generating and extracting tumor template DRRs from CyberKnife is time consuming, particularly for untrackable lesions, and limited our ability to create a larger data set. The images were split into three datasets for training, validation and testing (i.e., the holdout data set). To prevent potential bias, images in each dataset were kept completely separated by using a matlab function (splitEachLabel). The function took 216 images (80\%) for the training dataset, 27 images (10\%) for the validation dataset and 27 images (10\%) for the testing dataset.

    A transfer learning approach was used to train 5 different convolutional neural network architectures (AlexNet, DenseNet-201, Inception-ResNet-v2, VGG-19, and NASNet-Large, in order of increasing complexity defined by the number of GFLOPs, or decreasing efficiency) and to implement a binary classification of trackable and untrackable lung tumors. We tested five different architectures simply to compare results obtained with each network, and to show that the network choice does not influence the accuracy in predicting lung tumor trackability. These networks have been previously trained on over a million images from the ImageNet database. The original 5 convolutional neural network architectures were pre-trained on colored images with RGB channels, thus the grayscale 2D DRR images were saved as a three dimensional image with the same 2D grayscale DRR image repeated three times to mimic 3D RGB channels images. The DRR images have a size of 227x227 and they have been resized to fit the specific network requirements (e.g. Inception-ResNet-v2 requires input images of size 299x299x3). AlexNet was the simplest convolutional neural network that we tested. The AlexNet deep learning network architecture consists of 8 layers with learnable parameters \cite{Krizhevsky2017}.
	 
\begin{figure}[!htb] 
        \centering 
        \includegraphics[width=\figureWidth\linewidth]
        {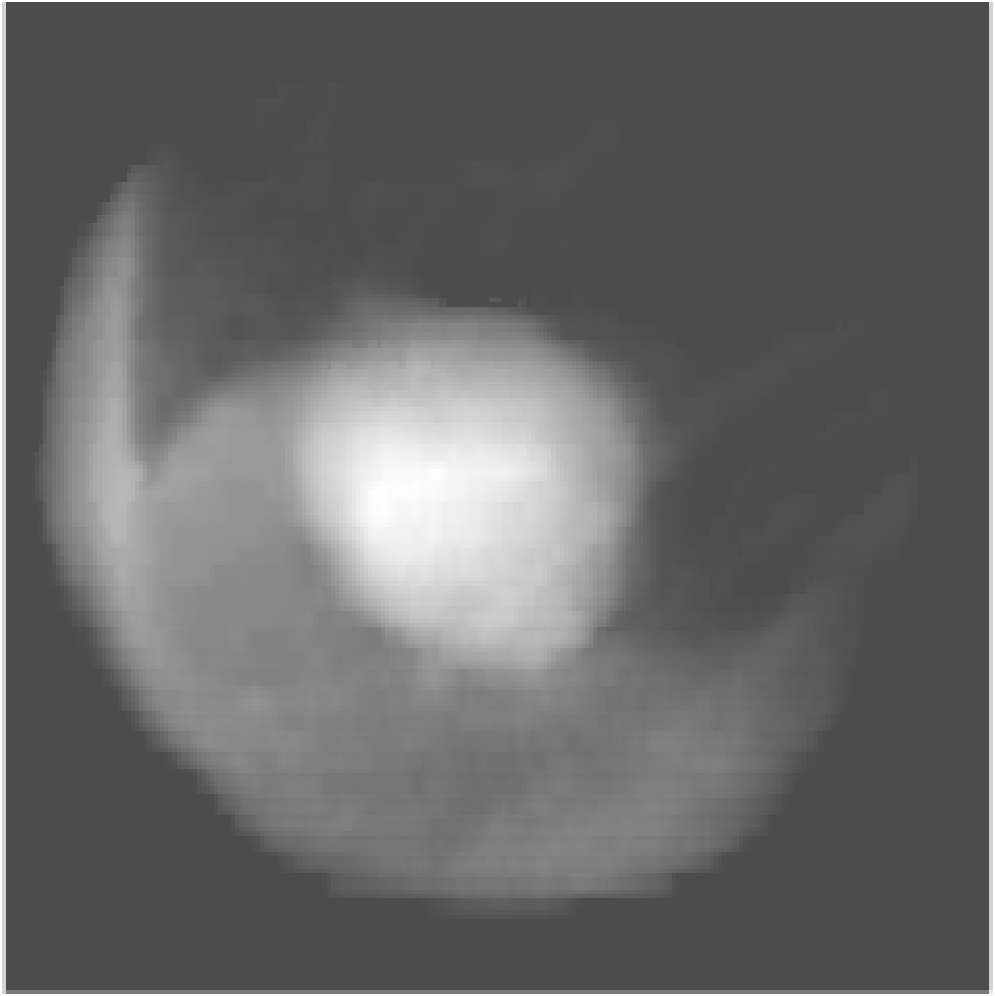}
        \caption[]{{Enhanced DRR image of a trackable lung tumor.}}
        \label{TrackableTumor}
    \end{figure}

\begin{figure}[!htb] 
        \centering 
        \includegraphics[width=\figureWidth\linewidth]
        {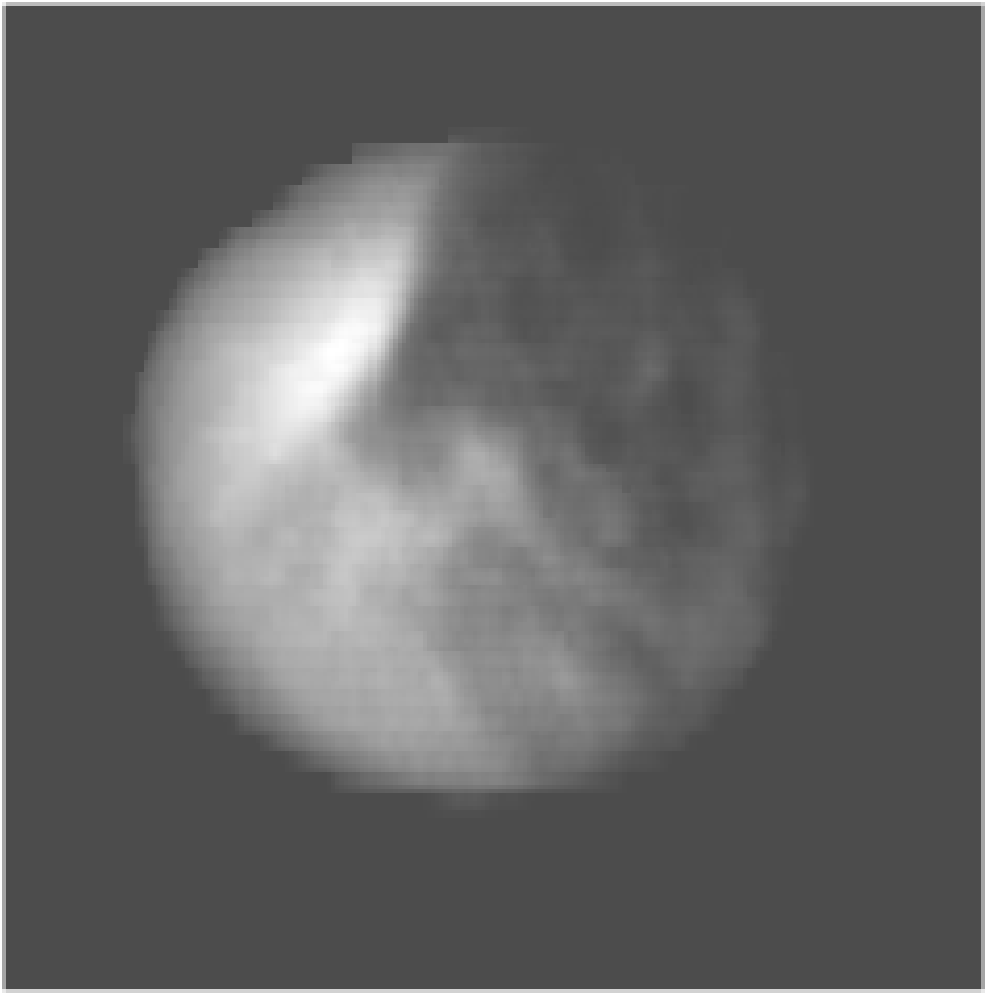}
        \caption[]{{Enhanced DRR image of a non-trackable lung tumor.}}
        \label{NonTrackableTumor}
    \end{figure}

    In the training process, the tumor template DRRs are classified as trackable or untrackable. A trackable tumor is shown in figure \ref{TrackableTumor}, and an untrackable tumor is shown in figure \ref{NonTrackableTumor}.  

\subsection{Deep-Learning Parameter Optimization} 
	
    The fine-tuning of a network with transfer learning is easier and faster than training a network from the ground up with randomly initialized weights. In order to perform a new classification task, the pretrained network has to be retrained, which requires to replace the last learnable layer and the final classification layer with new layers that are adapted to the new data set. The last fully connected layer of the pre-trained network was replaced with a new fully connected layer that has the same number of outputs as there are classes in the new data set. As this study concerns binary classification, there are 2 classes, trackable and untrackable tumors. The newly replaced last fully connected layer with learnable weights had a weight learning rate factor of 1e8, and a bias learning rate factor of 1e7.

    The retraining of the 5 different convolutional neural networks was performed by using an initial learning rate of 1e-7 for all the layers, except for the last fully connected layer which had a learning rate set to 10 so that the new classification task could be learned more easily and faster. The loss function hyperparameter of the classification layer calculates the cross-entropy loss for classification tasks with mutually exclusive classes. The mini batch size was set to 8, the validation frequency to 25, the maximal number of epochs to 20, the learn rate drop factor to 1, the learn rate drop period to 10000, the gradient threshold method was chosen as the global l2-norm, and finally shuffling was done at every epoch. The stochastic gradient descent with momentum (SGDM) solver was found to perform better than both the adaptive moment estimation (ADAM) solver and the root mean square propagation (RMSProp) solver to update the parameters of all 5 convolutional neural networks that were trained in this study (AlexNet, DenseNet-201, Inception-ResNet-v2, VGG-19, and NASNet-Large). No data augmentation was performed on the training images, validation images or testing images.
    
    \begin{figure*}
        \centering 
        \includegraphics[width=\figureWidth\linewidth]
        {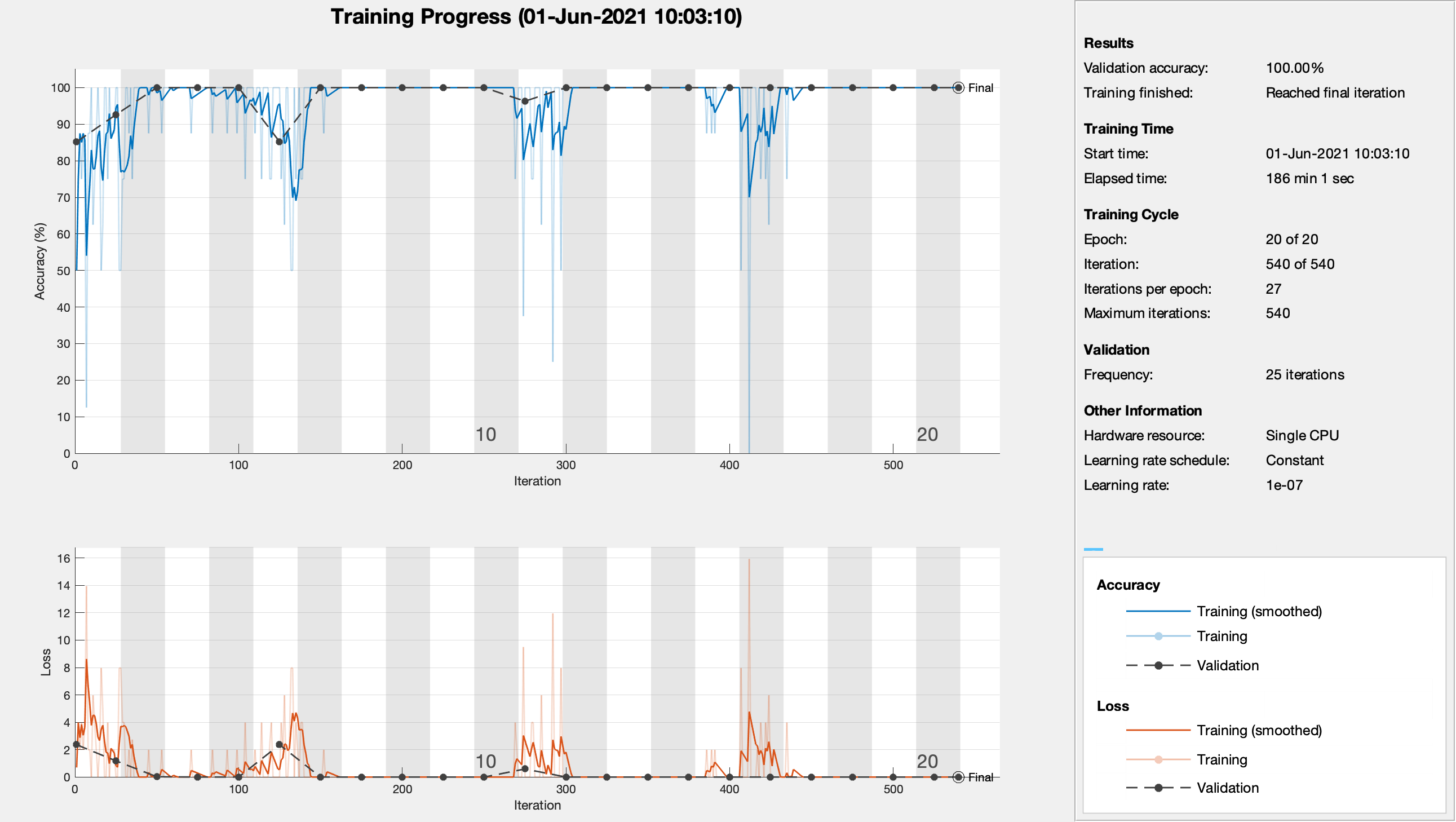}
        \caption[]{{The training progress and the final classification accuracy for the Inception-ResNet-v2 deep-learning network which results in 100\% validation accuracy in determining the feasibility to track the tumor.}}   
        \label{Deep_learning_model_training}
    \end{figure*}

    
\section{Results}

    \renewcommand{\arraystretch}{1.6}
    \begin{center}
    \begin{table*}
    \begin{tabular}{| p{3cm} || p{2cm} p{2cm} p{2cm} p{2cm} p{2cm} |}
    \hline
    Neural Network & AlexNet & DenseNet-201 & Inception-ResNet-v2 & VGG-19 & NASNet-Large \\ 
    \hline\hline
    Training Time (minutes) & 18 & 103 & 170 & 203 & 245 \\  
    \hline
    Classification Accuracy (\%) & 100 & 100 & 100 & 100 & 100 \\ \hline
    \end{tabular}
    \caption{Deep learning model training results.}
    \label{ResultTable}
    \end{table*}
    \end{center}
    
    The training time on a single 1.6GHz Intel i5 processor (no GPU) and 16GB of RAM was 18 minutes for AlexNet, 103 minutes for DenseNet-201, 170 minutes for Inception-ResNet-v2, 203 minutes for VGG-19, and 245 minutes for NASNet-Large.

    The training progress and the final classification accuracy for the Inception-ResNet-v2 deep-learning network is shown in figure \ref{Deep_learning_model_training}. The deep-learning based binary classification accuracy reached 100\% after training, both in the validation and the testing phase, for all 5 convolutional neural network architectures. There were no false classifications of trackable or untrackable lung lesions. The training results are shown in table \ref{ResultTable}.

\begin{figure} 
        \centering 
        \includegraphics[width=\figureWidth\linewidth]
        {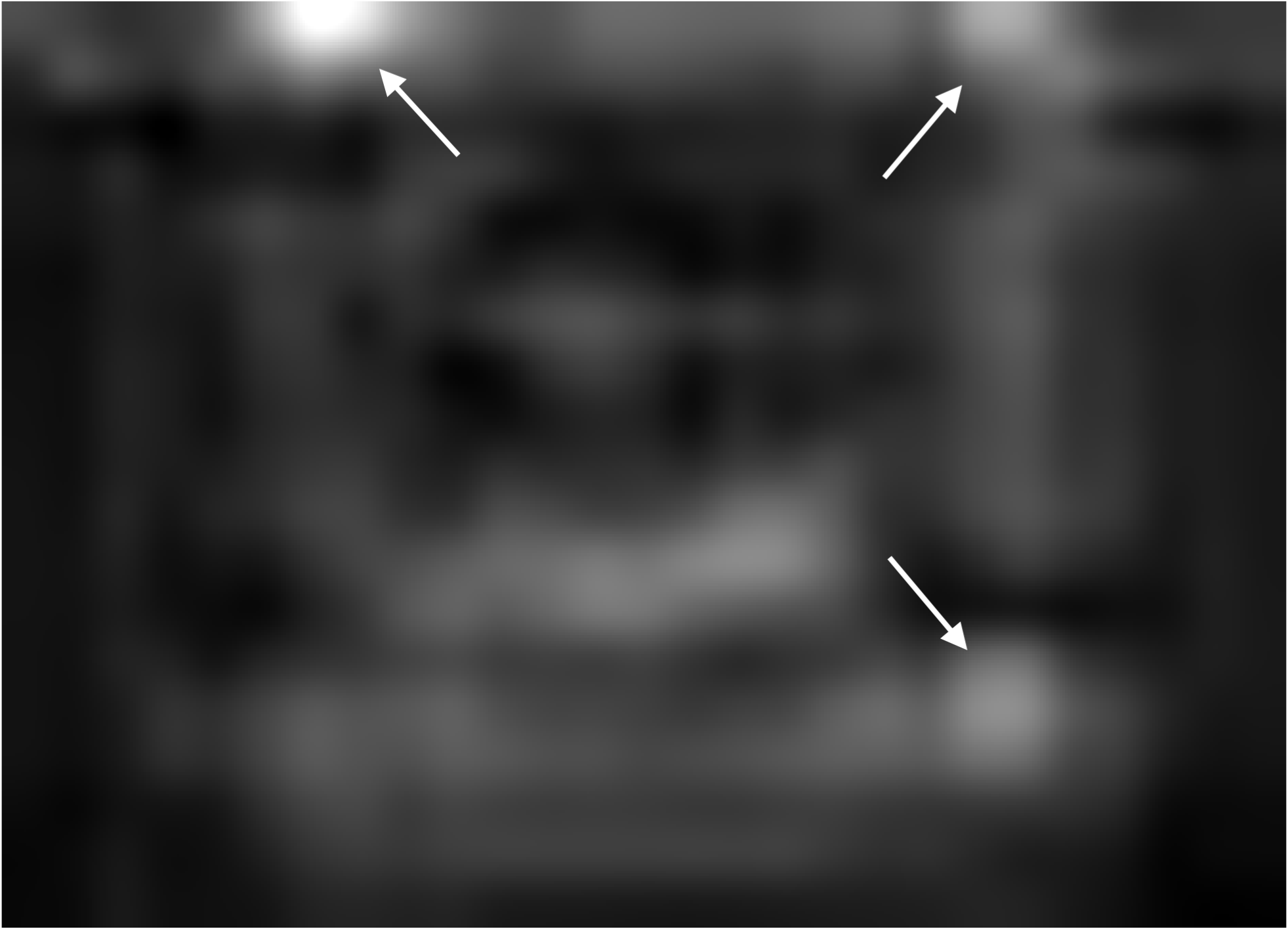}
        \caption[]{{Features extracted from the strongest activation channel in the middle layer of the Inception-ResNet-v2 network for a trackable lung tumor. Three full arrows are pointing to several bright white disks throughout the image, located at the corners of a blurry squared shape at the center of the image. These features are associated with a trackable lung tumor}.}
        \label{TrackableTumorFeatures}
    \end{figure}

\begin{figure} 
        \centering 
        \includegraphics[width=\figureWidth\linewidth]
        {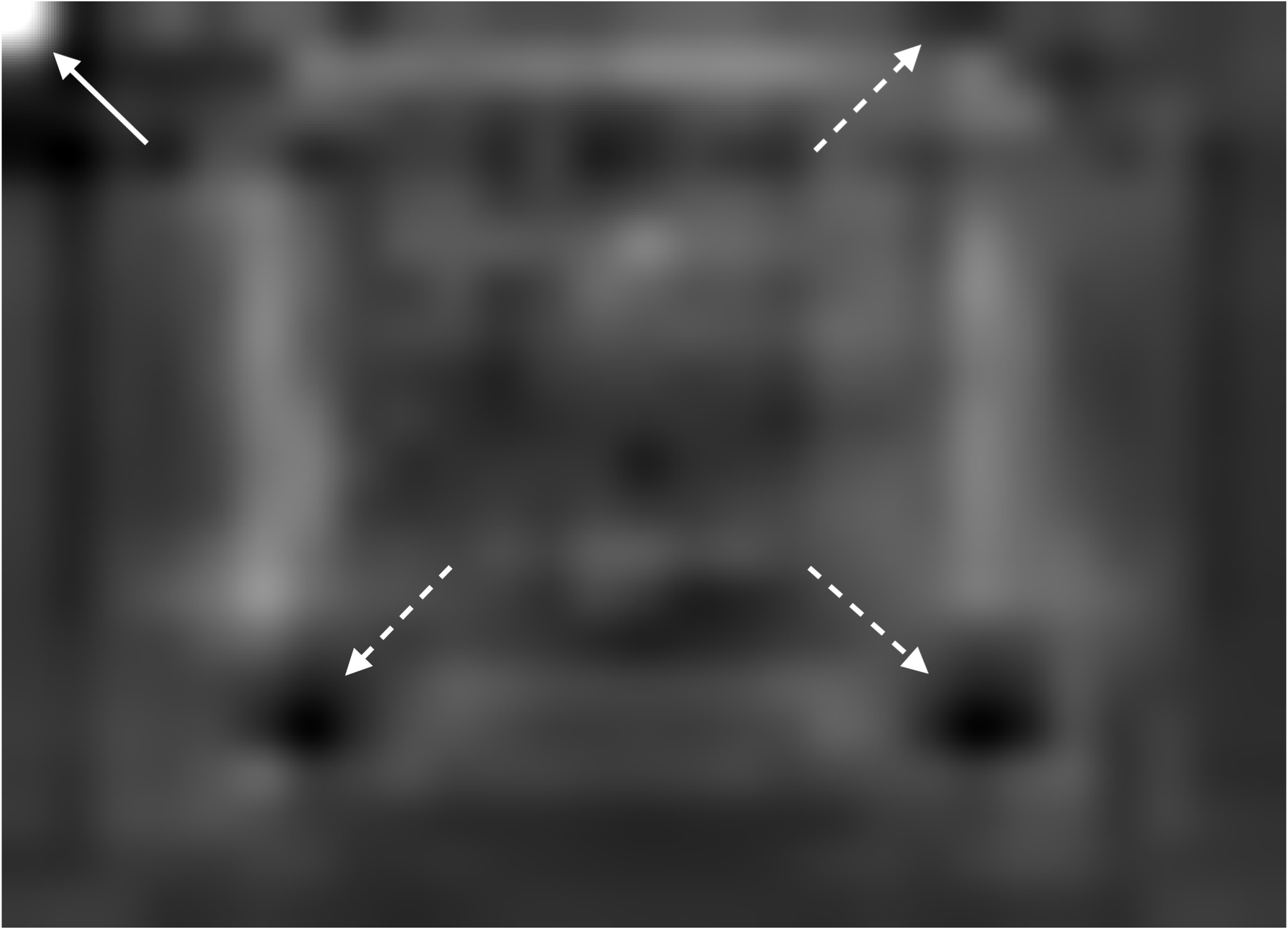}
        \caption[]{{Features extracted from the strongest activation channel in the middle layer of the Inception-ResNet-v2 network for a non-trackable lung tumor. One full arrow is pointing to a bright white disk on the top left corner of the image, whereas three dashed arrows are pointing to dark black disks which are located at the corners of a blurry squared shape at the center of the image. These features are associated with a non-trackable lung tumor}.}
        \label{NonTrackableTumorFeatures}
    \end{figure}

    The features extracted from the strongest activation channel in the middle layer of the Inception-ResNet-v2 network for a trackable tumor is shown in figure \ref{TrackableTumorFeatures}, and for a non-trackable tumor in figure \ref{NonTrackableTumorFeatures}. 
  
    These results show that this specific task of classifying the trackability of lung lesions based on tumor template DRRs is simple enough to be handled by an adequately tuned deep-learning model. The novel deep-learning based binary classification technique presented in this work can distinguish between trackable and untrackable lung lesions, and thus determine at the time of CT acquisition whether a patient could be treated with lung tumor tracking on CyberKnife, without ever having to perform an additional LOT simulation again. 
 
\section{Discussion}
		
    In this study, we presented a novel AI method leveraging a pretrained deep-learning convolutional neural network to distinguish features of trackable and untrackable lung lesions visible on DRR images. The trained network can be saved and used to classify the trackability of new DRR images generated from the CT scan at simulation time. This novel tool allows to automatically select the optimal tumor tracking method for patient undergoing lung SBRT on CyberKnife and forgo the time-consuming LOT simulation process, thus improving the workflow efficiency, saving the cost associated with resource utilization, and helping physicians in deciding whether a patient should undergo fiducial implantation or be treated on another machine.

    To our knowledge, this is the first study employing AI to predict successful lung tumor tracking candidates. Bahig et al. \cite{Bahig2013} investigated predictive factors for the applicability of lung tumor tracking in a cohort of 133 lung SBRT patients, pre-selected based on manufacturer recommendations and staff experience. They concluded that tumor size, volume, and density are most predictive of lung tumor tracking success. Lesions larger than 3.5 cm had a success probability of 80\%, but such probability decreased with decreasing tumor size. We chose to perform this study with tumor template DRRs, generated by the CyberKnife tracking system, to remove potential confounders in proving that determination of tumor trackability is feasible. Tumor template DRRs enhance the ability of the tracking algorithm to localize lesions, increasing the percentages of patients that can be treated with fiducial-free lung tumor tracking \cite{Jordan2010}.
    
    In the latest generation of lung tumor tracking algorithm developed in 2011, the registration between the live X-ray image and the tumor template DRR is performed by dividing the template in small square patches of a few millimeters, and by registering each patch independently. The registration for the whole template is calculated as the weighted sum of the individual patch registration, including only patches with a strong correlation with their location in the full content DRR \cite{Kilby2020}. Patches that include clearly identifiable features are given more importance. It is interesting to note how the strongest activation channel in the middle layer of the Inception-ResNet-v2 network extracts clearly identifiable features. As shown in figure \ref{TrackableTumorFeatures} 
    and figure \ref{NonTrackableTumorFeatures}, the detected patterns differ between trackable and untrackable images. In general for trackable lung tumor images (figure \ref{TrackableTumorFeatures}), there are three bright white circular regions throughout the image, located in the corners of a blurry squared shape at the center of the image. For most untrackable lung tumor images (figure \ref{NonTrackableTumorFeatures}), there is a bright white circular region that is located in the top left corner, and three dark black disks (instead of bright white disks) located in the corners of a blurry squared shape at the center of the image. It seems as if the algorithm can classify the images based on these features. Although these are typical patterns for most images, we do not have a definitive explanation as to the meaning of these patterns extracted by the algorithm which differ depending on lung tumor trackability. This highlights the black box nature of deep learning algorithms, but is potentially a step forward in trying to interpret the results. Machine learning and deep learning interpretability is an emerging field, and various interpretation tools have been proposed to understand how models make predictions \cite{LI2022}. Trying to understand the specific features found by our algorithm would require a separate study and is out of the scope of the present work.
 
    The results of this study would be implemented clinically by using one of our trained networks to feed it a pair of DRRs from a given patient, so that it can determine which, if any, of the images can be tracked. Although the analysis is performed independently for each image, the results are considered together in order to select the appropriate tracking method. If both images are deemed trackable, then we would proceed with 2-views tracking; if the algorithm deems that only one of the two images can be used for tracking, then we would proceed with 1-view tracking based on either image A or image B; if neither image can be used for tumor tracking, then spine tracking (0-view) or another treatment modality must be employed. Once a specific treatment pathway is chosen based on the algorithm's findings, the remaining aspects of the treatment process follow the normal clinical workflow.

    Although we have demonstrated the feasibility of using AI to predict the optimal tracking method, further work is needed prior to clinical implementation. This study has indeed three limitations: 1) small sample size - due to the time consuming process of extracting DRRs from the  CyberKnife system, our study included a limited number of images (n=230 trackable lesions, n=41 untrackable lesions). 2) DRR generation process - as described above, we elected to use tumor template DRR generated by the CyberKnife system. While tumor template DRRs enhance the tracking ability, full content DRRs are used to weight the importance of identifiable features, and for final visual evaluation of tracking results (the full content DRR is displayed at the control console and visually compared to the X-ray image). Although we were able to achieve 100\% classification accuracy, it is possible that including the full content DRR as a validation step could further improve the robustness of the method. Faster methods of DRR generation should also be considered. 3) Absence of internal and external cross-validation. While we have prevented bias by keeping the images in each set completely separated (i.e., we have used the holdout method by using an independent "testing" data set to perform an unbiased evaluation of our model, with the independent validation data set being used to tune our model and its hyperparameters), we have not employed an internal cross validation method (such as K-fold, leave-one-out, leave-p-out, etc.). Finally, our work represents a single institution retrospective study. An external validation using data from another institution, or prospectively analyzing images from patients undergoing LOT simulation, should be carried out to validate our findings.
    
    The method proposed in our study could be easily extended to lung cancer patients treated with other radiotherapy devices. Our deep learning algorithm should be able to classify any lung tumor as seen on a DRR as being trackable or untrackable, regardless of the specific technology used to deliver radiation, such as C-arm LINACs from various vendors and the Radixact system (Accuray Inc., CA, USA).


\section{Conclusion}
	
    In this work, we have demonstrated a new technique for leveraging AI and deploying a transfer learning approach for binary classification of tumor trackability for lung SBRT treatments on CyberKnife. We have evaluated 5 different convolutional neural network architectures to implement binary classification of trackable and untrackable lung tumors enabling an automatic selection of the optimal tumor tracking method for patient undergoing robotic radiosurgery. These 5 different deep-learning convolutional neural networks can distinguish features of trackable and untrackable DRR images generated from a CT scan at simulation time with 100\% accuracy in both the validation and testing phases, without false classifications. The chosen pretrained network can be saved and used to classify new DRR images generated from the CT scan at simulation time, thus enabling SBRT lung treatments to be performed on CyberKnife without ever having to complete a time-consuming LOT simulation again. This is an important tool for improving the efficiency of radiation therapy workflows on CyberKnife and saving precious resources and time in the clinic, both for patients and staff.


\section*{}
\addcontentsline{toc}{section}{\numberline{}References}
\vspace*{-20mm}










\bibliographystyle{./medphy.bst}    


\end{document}